\documentclass[slac_one]{revtex4}
\usepackage{graphicx}
\usepackage{fancyhdr}
\begin{document}

%Title of paper
\title{{\small{2005 ALCPG \& ILC Workshops - Snowmass,
U.S.A.}}\\ %% Please keep this conference title here
\vspace{12pt}
Overview and Charge - Snowmass Workshop 2005}

\author{Edmond L.\ Berger}
\email[e-mail: ]{berger@anl.gov}
\affiliation{Co-Chair, Organizing Committee, 2005 International Linear 
Collider Workshop} 
\affiliation{High Energy Physics Division, 
Argonne National Laboratory, Argonne, IL 60439}

%%%%%%%%%%%%%%%%%%%%%%%%%%%%%%%%%%%%%%%%%%%%%%%%%%%%%%%%%%%%%%%%%%%%%%%%%%%%%%
\begin{abstract}
This contribution to the published Proceedings records the opening talk 
I presented on the first morning of the 2005 International Linear 
Collider Workshop in Snowmass, CO, August 14 - 27, 2005.  It 
includes a summary of the motivation for the workshop, the scientific 
goals and charges for the working groups, the initial plans 
of the accelerator, detector, and physics groups, and the activities 
of the communication, education, and 
outreach group. This document also describes organizational 
aspects of the meeting, particularly the scientific committee 
structure, the self-organization of the working groups, the composition of 
the indispensable secretariat and computer support teams, and the sources 
of funding support. The report serves as an introduction to the 
proceedings whose individual papers and summary documents 
must be consulted for an appreciation of the accomplishments and 
progress made at Snowmass in 2005 toward the realization of 
an International Linear Collider.  

\end{abstract}

\maketitle

\thispagestyle{fancy}

%%%%%%%%%%%%%%%%%%%%%%%%%%%%%%
\section{INTRODUCTION}
\label{sec:intro}
Remarkable strides were taken in 2004 and 2005 toward the achievement 
of an international electron-positron linear collider having an initial 
center-of-mass energy of 500 GeV and the capability of 
extension to higher energy.  Particularly significant are the choice of 
superconducting technology by the International Technology Recommendation 
Panel in August 2004~\cite{techchoice}; the start of 
the Global Design Effort (GDE) led by Barry Barish; and the articulation of the 
essential and mutually supportive relationship of the Large Hadron Collider (LHC) 
and International Linear Collider (ILC) physics programs in the 2005 HEPAP subpanel 
document~\cite{hepapsubpanel}.  That great challenges lie ahead is an obvious 
understatement. A detailed design is required for the accelerator, along with 
full detector concepts, ever sharper physics arguments, and requisite funding.    

At the Linear Collider Workshop in Victoria in July/August 2004, consensus 
developed that the summer of 2005 would be an opportune time for the full community 
of physicists and engineers to gather for an extended period to work 
together to advance the design of the detectors and to understand better the 
scientific case for a linear collider. A proposal was then formulated in the 
American Linear Collider Physics Group (ALCPG), in consultation with 
international partners, to 
host a fully international detectors and physics workshop of duration long enough 
to facilitate substantial progress in addressing many of the challenges.  Uriel 
Nauenberg and I agreed to co-chair the Organizing Committee. In the 
fall of 2004, the ILC Steering Committee decided to hold the Second ILC 
Accelerator Workshop in conjunction with the Physics and Detector Workshop.  
This joint workshop was designed expressly with international participation in 
all the advisory committees and in the scientific program committees that 
organized the accelerator, detector, and physics activities.  

The Local Organizing Committee (LOC) selected Snowmass, Colorado, as the site of the 
workshop, formulated a set of specific goals, and wrote a 
four-part charge for the accelerator, detectors, physics, and education and 
outreach components of the workshop.  We submitted and defended successful 
funding proposals to the US Department of Energy and the National Science 
Foundation.  We made other funding appeals, primarily to national laboratories. 
Scientific committees were organized.  
A web site, http://alcpg2005.colorado.edu/, was developed and updated regularly.  
A computer support team and a secretariat 
were assembled. The scientific program was developed, special events were organized, 
wireless computer access was installed, meeting rooms were obtained and assigned, 
along with myriad other tasks.  

It is highly gratifying that over 670 enthusiastic participants are assembled in 
Snowmass, eager to contribute to the exciting endeavor that the ILC represents.

\section{CHARGE}
\label{sec:charge}      
There are four inter-related aspects of the Charge for the workshop.  

The primary ILC {\bf accelerator} goal is to define an ILC Baseline Configuration 
Document (BCD), to be completed by the end of 2005, and a research and 
development (R\&D) plan.  The BCD will be the basis for the design and costing 
effort as well as for developing the supporting R\&D program.
The accelerator groups at Snowmass will work toward 
agreement on the collider design, identify outstanding issues and develop 
paths toward their resolution, start documentation of the BCD, and identify 
critical R\&D topics and time scales. 

The {\bf detector} groups will develop design studies with a firm understanding 
of the technical details and physics performance of the candidate detector 
concepts, the required future R\&D, test-beam plans, machine-detector interface 
issues, beam-line instrumentation, cost estimates, and many related topics.  

On the {\bf physics} front, the goal is to advance and sharpen ILC physics 
studies, including precise calculations, synergy with the LHC, connections 
to cosmology and astrophysics, and, very importantly, relationships to the 
detector design studies.  

A fourth aspect of the charge has two components: 
to {\bf facilitate and strengthen the broad 
participation} of the scientific and engineering communities in ILC physics, 
detectors, and accelerators, and 
to {\bf engage the greater public} in the excitement of this work. 
The first component relates to the broad community of high energy physicists 
and engineers who may not have participated previously in linear collider 
activities or workshops.  The second addresses our fellow scientists in 
fields other than particle physics as well as members of the general 
public.  

\section{PLAN OF SCIENTIFIC ACTIVITIES}
\label{sec:plan}

\subsection{Accelerator}
The Working Groups established for the First ILC workshop at KEK in 2004 form the 
basis of the organizing units through Snowmass: Low-Emittance Transport and Beam 
Dynamics, Linac Design, Sources, Damping Rings, Beam Delivery, Superconducting 
Cavities and Couplers, Communications and Outreach.  In addition, six Global Groups 
were formed to work toward a realistic reference design: Parameters, Controls and 
Instrumentation, Operations and Availability, Civil and Siting, Cost and Engineering, 
and Options. 

The accelerator activities kick off on the first morning with a presentation 
by Barry Barish on the Global Design Effort.  The ILC working groups present 
introductory overviews in plenary sessions during the first afternoon. Lunchtime 
accelerator tutorials {\bf -- an accelerator school --} designed for experimenters 
and theorists begin on the second day of the workshop and continue through the 
next-to-last day.   

The very ambitious schedule of accelerator activities leads to plenary ILC 
Global Group summaries and working group summaries by the end of the first 
week of this two-week Snowmass workshop.  

\subsection{Detectors}

Three Detector Concept Studies are based on complementary philosophies: 
the Silicon Detector Concept (SiD), the Large Detector Concept (with 
Time Projection Chamber (TPC) tracking), and GLD (largest, with a TPC).  
These concepts are introduced in plenary sessions on the first afternoon.  
In the SiD approach, the calorimeter consists of a tungsten absorber and 
silicon detectors. A relatively small inner calorimeter radius of 1.3m is 
chosen.   Shower separation and momentum resolution are achieved with a 5 
Tesla (T) magnetic field and silicon detectors for charged particle tracking.
LDC, derived from the detector in the TESLA TDR, uses a somewhat 
larger radius of 1.7m, a silicon-tungsten calorimeter, and a large TPC for 
charged particle tracking.  GLD chooses a radius of 2.1m,  
a calorimeter with coarser segmentation, and gaseous tracking similar to LDC.  
The Snowmass workshop is a major opportunity for the collaborations to draft 
their Concept detector outline documents before the international 2006 
Linear Collider Workshop LCWS06 in Bangalore in March. 
 
Detector capabilities are challenged by the precise physics planned at the 
ILC.  The environment is relatively clean, but the detector performance must 
be a factor two to ten better than at LEP and the SLAC SLD.  Tracking, 
vertexing, calorimetry, software algorithms, and other aspects of the 
detectors are all on the agenda.
Among the leading questions and ambitions for the detector working groups are:
R\&D requirements; particle flow calorimetry for which a special session is 
organized on the second Monday; vertex detection at small radius; 
machine-detector interface issues (MDI) for which a joint 
MDI/accelerator/Concepts session is organized; agreement on feasible 
intersection-region (IR) parameters; the question of possibly two high 
energy IRs and detectors, discussed in a Town Meeting on Thursday afternoon 
during the first week; and 
the desirability of various options such as $e^+$ polarization, 
$\gamma \gamma$ collisions, $e \gamma$ collisions, and $e^- e^-$ collisions.  

\subsection{Physics}

The Physics activities begin on the first morning with presentations by 
Joe Lykken and Peter Zerwas on physics at the ILC and the LHC.  The ILC 
offers the capability to control the collision energy, polarize one or 
both beams, and measure cleanly the particles produced.  It will allow 
us to zero in on crucial features of the physics landscape, a rich world  
of Higgs bosons, supersymmetric particles, dark matter, and extra spatial 
dimensions. Four physics 
working groups are assembled under the headings Higgs, Supersymmetry (SUSY), 
Beyond the Standard Model (BSM), and top quark plus quantum chromodynamics 
(Top/QCD), along with three cross-cutting Special Topics groups: Cosmology 
Connections, LHC/ILC Connections, and precise high-order calculations (Loopfest).  
Plenary sessions for the Physics working groups take place on the 
first Tuesday at which the conveners outline the activities planned for 
each group.  A two-day Loopfest Conference takes place Thursday and Friday 
of the first week, and a day is devoted to Cosmology and the ILC on the 
second Wednesday, August 24. 

A partial menu of physics topics includes Physics benchmarks, with a 
plenary session on the first Tuesday afternoon; the Higgs mechanism --  
lessons about electroweak symmetry breaking at the ILC that we can learn no 
other way; SUSY -- determination of masses and other 
parameters at the focus point and for other Snowmass points and slope 
scenarios from a combination of LHC and ILC data; extra-dimensions and strings;  
and precise high-order calculations to match the expected high precision of the 
ILC data. Permeating all these discussions is the paramount question of what    
ILC detector capabilities are needed.  Detector benchmarking 
adds an important dimension and emphasis to the physics discussions at Snowmass. 

There is a tremendous amount to accomplish in the two weeks of this workshop 
leading to the physics and detector summary talks at the end of the second 
week. 

The individual papers and summary documents in these Proceedings  
must be consulted for an appreciation of the accomplishments of the 
workshop and progress made at Snowmass in 2005.  A summary written 
for a more general audience is published in the December 2005 issue of 
the {\em CERN Courier}~\cite{Courier}.  
 
\section{COMMUNICATIONS, EDUCATION, and OUTREACH}

One of our important responsibilities is to engage our fellow citizens 
in the excitement of particle physics.  Some of what we do may be mysterious,  
but none of it is a secret. Most of those with whom we speak are 
more curious and genuinely interested than we might guess.  

A {\bf Dark Matter Cafe and Quantum Universe Exhibit} 
will be set up on the Snowmass Mall during our first weekend here, with volunteers 
among us enjoying conversations with local residents and tourists.  A 
{\bf Workshop on Dark Matter and Cosmic Ray Showers} is organized for high 
school teachers on Friday of this first week, followed by a display of working 
cosmic ray shower detectors on the Aspen Mall on Saturday, staffed by members of 
the Education and Outreach working group and other volunteers from 
among the participants.  A {\bf Physics Fiesta} takes place on Sunday at the 
Roaring Fork High School in Carbondale, and volunteers, especially those with 
Spanish language skills, will meet with students, family members, and teachers.   

Two Public Lectures are scheduled at 6:30 PM, one in Aspen on August 17 by 
Young-Kee Kim entitled {\em $E = mc^2$: Opening Windows on the World}, and the 
second in Snowmass on August 22 by Hitoshi Murayama,  
{\em Seeing the Invisible -- Challenge to 21st Century Particle Physics and Cosmology}.  

Complementing these activities organized by the Education and Outreach Committee of the 
Physics and Detectors component of this overall joint workshop, the ILC Communications 
group is hosting a series of workshops and invites all participants.  The new website 
http://www.linearcollider.org is being launched at Snowmass, starting with daily coverage of 
the workshop, along with ILC NewsLine, http://www.linearcollider.org/newsline/, a weekly 
newsletter free to all subscribers. 

\section{OTHER SPECIAL EVENTS}

The importance of involving industry in the design and execution of the 
accelerator and detectors is recognized in an ILC Industry Forum on Tuesday, 
August 16 at 7:30PM.  

The ILC International Steering Committee (ILCSC) convenes in an all day meeting on 
Tuesday August 23.

An evening Forum on Tuesday the 23rd addresses 
{\bf Challenges for Realizing the ILC: Funding, Regionalism, and International Collaboration}.  
Eight distinguished speakers representing committees and funding agencies with direct 
responsibility for the ILC share their wisdom and perspectives:  Jonathan Dorfan (ICFA 
Chairman), Fred Gilman (US-HEPAP), Pat Looney (formerly of OSTP), Robin Staffin (US-DOE), 
Michael Turner (US-NSF), Shin-ichi Kurokawa (ACFA Chair and incoming ILCSC Chair), Roberto 
Petronzio (Funding Agencies for the Linear Collider, FALC), and Albrecht Wagner (incoming ICFA 
Chair).  Ample time is set aside for animated questions and comments from members of 
the audience.    

Workshop Dinners take place on Thursday of the first week and Wednesday of the second week.  
Tickets are available for purchase.

\section{FUNDING SUPPORT}    

We acknowledge and are most appreciative for generous grants from the US 
Department of Energy (DOE) and the US National Science Foundation (NSF).  The DOE 
funds underwrite workshop expenses, such as meeting rooms and secretariat 
and computing expenses.  The NSF funds provide subsidies for the local 
expenses of young participants and for the education and outreach activities.  
 
We received indispensable financial contributions from Argonne National 
Laboratory, Cornell Laboratory for Elementary Particle Physics, 
Brookhaven National Laboratory, Lawrence Berkeley National 
Laboratory, and Thomas Jefferson Laboratory to assist with workshop expenses.  We 
are most grateful for this support and for grants from 
the Universities Research Association (URA) and from Stanford University that 
subsidize the opening reception on Sunday, August 14, and the two workshop 
dinners.  
                        
Support for participants from Europe was received from DESY, PPARC (UK), 
IN2P3 (France), and CERN.  We acknowledge both the funds provided and the 
instrumental assistance in obtaining and administering these funds from  
Rolf-Dieter Heuer, David Miller, Francois Richard, and Jos Engelen. 

The workshop could not have taken place without the enormous in-kind contributions 
from Fermilab (leadership of the secretariat, members of the secretariat and 
computer support teams, and equipment), SLAC (proceedings, and members of the secretariat 
and computer support team), and Cornell (leadership of the computer support team).  

\section{HEAVY LIFTING} 

\subsection{Local Organizing Committee}

Recognition of the tremendous contributions of talent and time from many individuals
begins with the members of the Local Organizing Committee.  Foremost among these  
is my co-Chair Uriel Nauenberg, along with his associates Valerie Melendez and 
webmaster Will Ruddick from the University of Colorado.  
ALCPG Co-Chairs Jim Brau (Oregon) and Mark Oreglia (Chicago) along with accelerator 
community representatives Shekhar Mishra (Fermilab) and Nan Phinney (SLAC) round out 
the LOC.  I am indebted to all my colleagues on the LOC for their 
constant dedication and many hours of effort on behalf of this workshop 
during the past year. And we have much left to do!    

\subsection{Executive Committee}

Members of the international executive committee provided timely and valued assistance 
in many respects.  They are Barry Barish (Caltech), Edmond Berger (Argonne, co-Chair), 
James Brau (Oregon), Sally Dawson (Brookhaven), Rolf Heuer (DESY), David Miller (UC London), 
Shekhar Mishra (Fermilab), Uriel Nauenberg (Colorado,co-Chair), Mark Oreglia (Chicago), 
Hwanbae Park (Kyungpook National University), Michael Peskin (SLAC), Tor Raubenheimer (SLAC), and 
Hitoshi Yamamoto (KEK).   

Maury Tigner of Cornell and Pier Oddone of Fermilab were generous advisors, 
sources of good counsel and {\em savoir faire}.  

\subsection{Working Group Organizing Committees}

Four working committees were appointed by the Local Organizing Committee and 
asked to define the topics of the working groups and to secure  
conveners of these working groups.  The members of these four committees for 
the accelerator, detector, physics, and education and outreach efforts are: 
\subsubsection{Accelerator} 
David Burke (SLAC), Jean-Pierre Delahaye (CERN), Gerald Dugan (Cornell), 
Hitoshi Hayano (KEK), Steve Holmes (Fermilab), Olivier Napoly (Saclay), 
Kenji Saito (KEK), Nick Walker (DESY), and Kaoru Yokoya (KEK).
\subsubsection{Detectors} 
Philip Bambade (Orsay), Ties Behnke (DESY), Tiziano Camporesi (CERN), 
John Jaros (SLAC), Dean Karlen (Victoria), Akiya Miyamoto (KEK), 
Mark Oreglia (Chicago, Chair), Daniel Peterson (Cornell), 
Harry Weerts (Michigan State/Argonne), and Satoru Yamashita (Tokyo).
\subsubsection{Physics}
Sally Dawson (Brookhaven, vice-Chair), Jonathan Feng (UC Irvine), Rohini 
Godbole (Bangalore), Norman Graf (SLAC), Howard Haber (UC Santa Cruz), 
Kaoru Hagiwara (KEK), Joseph Lykken (Fermilab), Michael 
Peskin (SLAC, Chair), W. James Stirling (Durham), 
Rick Van Kooten (Indiana), and Peter Zerwas (DESY).
\subsubsection{Education and Outreach}
Marjorie Bardeen (Fermilab), Neil Calder (SLAC), 
Ulrich Heintz (Boston University), Judy Jackson (Fermilab), 
Hitoshi Murayama (UC Berkeley, Chair), 
and Gregory Snow (Nebraska).  

\subsection{Working Groups}

In a sense, the two-dozen or so working groups are self-organizing units. 
Individuals choose whether to participate, and the groups define 
how they will best use their time and talents at Snowmass to achieve 
their goals.  Nevertheless, the international conveners of the working 
groups, with representation 
from all regions, are the unsung heroes and heroines.  They must 
cajole and motivate their members, keeping them focused.  I apologize 
for not listing all of the conveners by name. Their identities and the 
agendas of the working group programs can be found from links on the 
workshop program page:  
http://alcpg2005.colorado.edu:8080/alcpg2005/program/. 

The number of different working groups requires 22 
meeting rooms for break-out sessions.  A few of these are obtained by 
partitioning the large ballroom used for the plenary sessions on 
the first day and for the summary talks on the final two days. Many 
of the other meeting rooms are distributed among the different 
properties at Snowmass.  All are within walking distance of 
the central Conference Center.  A spreadsheet on the workshop 
web page lists room assignments, but updates, additions, changes will be 
made as needed.  Mark Oreglia did a heroic job juggling a complex mix 
of competing room requirements!  Rooms are equipped with LCD projectors 
and screens; overhead projectors are also available by special 
request.  Working group conveners are asked to provide laptop 
computers for projecting presentations, with file transfer to be done via 
USB thumb drives.  

\section{SECRETARIAT, COMPUTER TEAM, and SNOWMASS PERSONNEL}

\subsection{Snowmass Personnel}

The particle physics community in the United States has run many extended 
summer workshops in Snowmass over the past two decades.  We return  
because we find the concentrated facilities we need for a large gathering  
with many breakout rooms, and a group of local representatives who go the 
extra mile to provide what a group of physicists requires.  I wish to acknowledge 
the forthcoming assistance received from Jim Pilcher and his staff at the 
Silvertree Hotel, Jim O'Leary of the Snowmass Village Resort Association, 
and Maidy Reside of Top of the Village.  

\subsection{Secretariat} 

The local staff at Snowmass assisted admirably with lodging reservations 
for participants, but the innumerable aspects of registrations, negotiations 
with vendors, and daily logistics have fallen on the shoulders of the dedicated 
and most capable Secretariat headed by Cynthia Sazama of Fermilab.  Cynthia was 
an enthusiastic member of the organization from the first moment she was 
approached and asked to participate.  Her team is made up of Maura Chatwell, 
Albe Larsen, and Naomi Nagahashi (SLAC) and Carol Angarola, Jody Federwitz, 
and Suzanne Weber (Fermilab). 
Their main office is at the Top of the Village (ToV) in condo unit Slope 210, 
with weekday hours of operation 8:00 AM - 6:00 PM, and Saturday hours 
8:00 AM - 1 PM.  Registration takes place in the Conference Center on the 
first day and in ToV Slope 210 thereafter.  

\subsection{Computer Facilities and Support}

Chips were down many months ago when we desperately needed a person 
to define the configuration of the workshop computer facility 
and to make it work.  Maury Tigner stepped up as he has done 
so many times in his distinguished career. Maury offered the services of 
Ray Helmke, Director of the Computing Facility at Cornell's 
Laboratory for Elementary Particle Physics.  The computer support team 
assembled under Ray's superb leadership includes his 
deputy John Urish (Fermilab), David Tang and Quinton Healy (Fermilab), 
Mike DiSalvo and Ken Zhou (SLAC), Bryan Abshier (LBL), and Andrew Hahn, Joseph 
Proulx, Martin Nagel, and Jason Gray (U Colorado).  

Part of the networking and computing capacity is satisfied by an existing 
T1 line that 
serves the Silvertree Hotel and the Snowmass Conference Center, but a 
second dedicated line was brought into the Conference Center specifically 
for the greater requirements of the workshop.  

The computer setup for this workshop is principally a laptop based facility, 
relying on equipment brought by individual participants, most functioning in 
wireless mode.  There are also some hardwired connections and 
printers in the computer 
center rooms located in three condos at Top of the Village (ToV), Trails 
109, 108, and 105, where 18 Windows PC's on loan from Fermilab may also be 
found. The computer center is staffed daily during the hours 8 AM to 10 PM.   
Wireless access in the computer center and in condos in the ToV complex 
relies on a commercial cable system recently installed by ToV.  

The computer support team set up the facility in the three condos in ToV
during the week before the workshop and tested it.  They prepared a 
descriptive document, available on the workshop web site.  

Our workshop and the international ILC effort owe a debt of 
gratitude to all members of the Computer Support Team and of the Secretariat.  

\section{PROCEEDINGS}

Electronic files of all presentations at the workshop are 
linked on the web pages. Proceedings will appear on the SLAC Electronic 
Conference Proceedings Archive (eConf), a permanent repository for conference 
proceedings, and a CD will be produced of written 
contributions.  Norman Graf (SLAC) is leading the editorial effort.  The 
Proceedings link on the front page of the web site 
may be consulted for instructions and page limits.  

\section{L'ENVOI}

The organization of this workshop has benefited from the dedication and 
talents of scientists, engineers, and support personnel from many 
institutions and all regions of the world.  
I reviewed the motivation for the workshop, the scientific goals and 
charge to the working groups, the initial plans of 
the accelerator, detector, and physics working groups, and the 
activities of the education and outreach working group. I also summarized 
organizational aspects of the meeting, particularly the scientific 
committee structure, the self-organization of the working groups, the 
composition of the indispensable secretariat and computer support teams, 
and the sources of funding support. 
 
This 2005 Snowmass Workshop is now in your capable hands to:
%\addtolength{\itemsep}{-0.5\baselineskip}
\begin{itemize}
\addtolength{\itemsep}{-0.5\baselineskip}
\item design the accelerator,
\item flesh out the detectors,
\item hone the physics reasoning, and
\item engage your fellow citizens
\end{itemize}
all within a very ambitious time scale.  

\begin{acknowledgments}
Work in the Argonne High Energy Physics Division is supported by 
the US Department of Energy, Division of High Energy Physics, 
under contract W-31-109-ENG-38.

\end{acknowledgments}

\end{document}